\documentclass{llncs}
\usepackage{lncsedoc}
\usepackage[final]{graphicx}
\begin{document}
\title{Are Local Causal Models of Quantum Theory Feasible at All?}

\author{ Hans H. Diel}

\institute{Diel Software Beratung und Entwicklung, Seestr.102, 71067 Sindelfingen, Germany,
diel@netic.de}
\maketitle

\begin{abstract}
This article presents an analysis of the extent to which local causal models or local realistic models of quantum theory (QT), including
quantum field theory (QFT), are theoretically possible and practically feasible in light of the present state of these theories. Quantum physicists
consider Bells famous inequality and its violation in experiments to be a strong indication that local realistic or local causal models of QT are
not possible and that quantum theory as a whole is therefore not a local realistic or local causal theory. Based on a proposed definition of
a "formal causal model" for a theory of physics (such as QT), this paper investigates the possibility of having a local causal model for QT. Areas of QT are
identified in which the construction of a causal model is impeded because of deficiencies in the state of the respective theory. It is shown that the
removal of the deficiencies can be achieved by the provision of a causal model. Whereas the construction of a causal model of QT, including QFT,
appears to be feasible after the removal of certain deficiencies, the construction of a local (causal) model does not appear to be possible. As a consequence of the
conclusion that local (causal) models of QT/QFT are not possible, if a strong interpretation of locality is assumed, a locality model is proposed in which the non-localities are confined to "quantum objects". 
\end{abstract}


\section{Introduction}

The origin of the discussion on the possibility of local causal models of quantum theory (QT) dates back to A. Einstein, who questioned the completeness of QT
(although he did not express concerns regarding the possibility of a local causal model of QT). Later, Einstein, Podolsky and Rosen made these arguments more precise in the form
of the famous EPR Gedanken experiment 
 (see \cite{EPR}). J. Bell supported the doubts about the completeness and/or correctness of QT by the formulation of the famous Bell's inequalities  (see \cite{Bell}). 
Based on assumptions that are considered to be "obtained from ideas about objective reality and causality that appear
to be good common sense" (phrasing used in  \cite{Jordan} , p. 125), Bell's inequalities request measurement results for the EPR-experiment that are in conflict with the predictions of QT.
Years after the publication of Bell's inequalities, Aspect et al. succeeded in performing real EPR-experiments (see \cite{Aspect}). The results of the
experiments agreed with the predictions of QT and violated Bell's inequalities. Because the establishment of Bell's inequalities was based on ideas about
objective reality and causality that appear to be good common sense, the violation of the inequalities was a surprise to many physicists. The discussion regarding the
implications of the violation of the inequalities subsequently began and has not yet come to an end.

The most frequent interpretation of the violation of Bell's inequalities in experiments is that the establishment of the inequalities by J. Bell is a clear
indication (or even proof) that local realistic or local causal models of QT are not possible. Follow-up discussions have centered around the subjects "completeness of QT", "realistic model of QT", "causal model of QT" and "locality in QT". Discussions have presented alleged proofs (e.g., for the completeness of QT), as well as
rejections of the proofs. The terms that have been discussed (e.g., completeness, realistic, local) were used with respect to the meanings that they were presumed to have in their original
usages. Attempts to define the terms more precisely were, in general, driven by philosophical and ontological considerations.

This paper joins the discussion, with an emphasis on the possibility of a causal model of QT. The subjects "local model of QT" and "realistic model of QT"
will also be addressed, based only on the causal models. The first part of the article (Sections 2 - 4) addresses the causal models. Instead of attempting to
achieve the (necessary) clarification of key terms (e.g., causal model, realistic model, local model) by a deeper ontological analysis, the paper begins in Section 2 with proposals for formal definitions of the above terms and types of models. It is not possible to present a complete formal treatment of
the subject (which would be too lengthy and would deviate from the essential points) within this paper. After the establishment of precise definitions of the
different types of models, it may be possible to investigate to what extent these terms are applicable to QT. For example, the following questions should be considered:
\begin{enumerate}
\item Are local causal models of QT feasible?
\item Is QT complete?
\item Are local realistic models of QT feasible?  
\end{enumerate} 
Based on an analysis of these questions, it appears that there are areas within QT in which the analysis is impeded (or even impossible) because the respective area
is not (yet) described in a way that can be translated into a formalism such as a causal model. It is shown here that the removal of deficiencies is a prerequisite
for the construction of a causal model and that the removal of deficiencies can, in fact, be achieved by the provision of a suitable causal model.
The problematic QT areas (e.g., the measurement problem, "interference collapse rule", entanglement and QFT interactions) are described in Section 3.

In Section 4, an overall causal model of QT/QFT is proposed, which implies proposals for the solutions in the problem areas that are described in Section 3.
The proposed causal model implies realism, but it is not a local model according to the definition of locality given in Section 2.

Locality is discussed (starting in Section 5) based on the causal model described in Section 4. In Section 5, a definition of a "local causal model" is given,
and a discussion is provided regarding the extent to which the causal model described in Section 4 may be called a local causal model. The conclusion is that the causal models
of the problem areas contain non-localities if the (narrow) definition of locality (as given in Section 5) is applied. As a consequence, in Sections 6 and
7, a refined causal model is described in which the non-localities are confined to "quantum objects".

\section{Models of theories of physics}
For a meaningful discussion on the feasibility of a causal  models of QT, it is necessary to discuss first what constitutes a model of an area of physics. A most proper discussion can be achieved if it is possible to identify a 
\emph{formal} model of a theory of physics. 
Formal specifications are typically given in terms of mathematics. For a formal specification of a causal model, the typical mathematical equations in terms of differential equations, matrices, tensors, etc., are not sufficient. For the specification of physical processes and state transitions, it must be possible in addition to specify algorithms, case distinctions, iterations, and the structure of compound physical objects.

In  \cite{DielComplete}, a "formal causal model" of a theory of physics is described, which is also a suitable starting point for our discussion of local causal models of QT/QFT in the present article.

\subsection{Formal causal model of a theory of physics}

The specification of the formal model of a theory of physics consists of (1) the specification of the system state and (2) the specification of the laws of physics that define the possible state transitions when applied to the system state. For the formal definition of a causal model of a physical theory, the laws of physics are represented by a "physics-engine". The physics-engine acts upon the state of the physical system. The physics-engine continuously determines new states in uniform time steps. For the formal definition of a causal model of a physical theory, the continuous repeated invocation of the physics-engine to realize the progression of the state of the system is assumed.
\\
\\
$ systemstate := \{ spacetimepoint ... \} \\
\; \;   spacetimepoint :=\{  t,  x_{1}, x_{2}, x_{3}, \psi \} \\
\; \;  \psi :=  \{ stateParameter_{1}, ... , stateParameter_{n} \} $
\\
\\
$ systemEvolution( system\; \; S ) := \{    \\
 S.t = 0; S.x_{1}=0; S.x_{2}=0; S.x_{3}=0;  \\
 S.\psi = initialState; \\
 \Delta t = timestep;     ||\;  must\; be\; positiv  \\
 DO \;  UNTIL ( nonContinueState( S ) )  \{  $

   physics-engine $(  S, \Delta t );  $ \\
\}  
\\
\\
physics-engine $(  S,  \Delta t ) := \{  $

$ \; \;    S = applyLawsOfPhysics( S, \Delta t );  $  \\
\}  
\\
\\
The refinement of the statement $  S = applyLawsOfPhysics( S, \Delta t ); $ defines how an "in" state s  evolves into an "out" state s.

    $      L_{1} : \: IF \: c_{1}(s)\: THEN \: s = f_{1}(s); $

    $      L_{2} : \: IF \:c_{2}(s) \: THEN  \: s = f_{2}(s); $

            ...

      $     L_{n} : \: IF \:c_{n}(s) \: THEN   \:s = f_{n}(s); $
\\
The "in" conditions $ c_{i}(s) $  specify the applicability of the state transition function 
$ f_{i}(s) $ in basic formal (e.g., mathematical ) terms or refer to complex conditions that then have to be refined within the formal definition.

The state transition function $ f_{i}(s) $ specifies the update of state s in basic formal (e.g., mathematical) terms or refers to complex functions that then have to be refined within the formal definition.
\\
To enable non-deterministic theories ("causal" does not imply deterministic) an elementary function 

RANDOM(valuerange, probabilitydistribution)
\\
may also be used for the specification of a state transition function. 
\\ 
The set of laws $ L_{1}, ... ,L_{n} $ has to be complete, consistent and reality conformal (see \cite{DielComplete}  for more details).

\subsubsection{Example1 - A causal model:}

Many areas of physics can be described by starting with a specific Lagrangian. For a description of the causal relationships, i.e., the evolution of the system state, the equation of motion is the 
major law. The equation of motion can be derived from the Lagrangian by using the Euler-Lagrange equation.
\\
The Lagrangian for classical mechanics is
\\ L = V - T with 
\\
$ V=V(x), T = \frac{1}{2} m \dot{x}^{2} $. 
\\
The Euler-Lagrange equation leads to the equation of motion

 $     m \ddot{x} =  \frac{\delta V}{\delta x} $.
\\
The specification of the laws of classical mechanics can be given by a list ($ L_{1}, ... ,L_{n} $ ) that distinguishes different cases or by a single general law. The single general law is
\\
  $  L_{1} $ :  IF  ( TRUE ) THEN   FOR (all Particles $ P_{i} $ ) \{
\\
  $   \;\;  P_{i} = applyEquationOfMotion(P_{i}); $ \}
\\
Thus, the system state has to contain
\\
$ systemstate := \{ $

   $  space $ ; 

   $  particles =   P_{1}, ...,  P_{n} $;

   $  field \;\;  V = V(x)  $;

   $  Particle \;\;   P = \{m, x, \ddot{x}, \dot{x} \} $ 
\\
\}

\subsection{Spatial causal model}

A causal model of a theory of physics is called a \emph{spatial} causal model if 
(1) the system state contains a component which represents a space, and (2) all other components of the system state can be mapped to the space.

There exist numerous textbooks on physics (mostly in the context of Relativity theory) and on mathematics which define the essential features of a "space". For the purpose of the present article a more detailed discussion is not required. For the purpose of this article and the subject locality it is sufficient to request that the space (assumed with a spatial model)
supports the notions of position, coordinates, distance, and neighborhood. 

\subsubsection{Example2 - A spatial causal model:}

A possible type  of a spatial causal model is the cellular automaton (CA). 
The classical CA consists of a k-dimensional grid of cells. The state of the CA is given by the totality  of the states of the individual cells.

    state $ =  \{ s_{1}, ... , s_{n} \} $
\\
With traditional standard CAs, the cell states uniformly consist of the same state components

   $    s_{i} =  \{ s^{1}_{i}, ... , s^{j}_{i} \} $
\\
Typically, the number of state components, j, is 1, and the possible values are restricted to integer numbers.
The dynamical evolution of the CA is given by the "update-function", which computes the new 
state of a cell and of the neighbor cells as a function of  the current cell state.
\small{
\begin{verbatim}
Standard-CellularAutomaton(initial-state)  :=     // transition function
state = initial-state;
DO FOREVER {
   state = update-function(state, timestep);
   IF ( termination-state(state))  STOP;
}
\end{verbatim}
}
\normalsize
The full functionality and complexity of a particular CA  is concentrated in the update-function. As Wolfram (see \cite{Wolfram}) and others (see, e.g., \cite{Ilachinski}) have shown, a large variety of process types (e.g., stable, chaotic, pseudo-random, and oscillating) can be achieved with relatively simple update-functions.

\subsection{Realistic model}

When in the QT literature the term "realistic model" is used in connection with entanglement and Bell's inequality,
different variants can be found for the definition of "realistic model".
In \cite{Laudisa2}  F.  Laudisa described
"realism" as 
"a condition which is often formulated, even recently, as the
idea that physical systems are endowed with certain pre-existing properties, namely
properties possessed by the systems prior and independently of any measurement
interaction and that determine or may contribute to determine the measurement outcomes ".
\\
This definition refers to "measurement interaction" and "measurement outcome" and requests a causal relationship between the system state being measured (called "preexisting properties") and the "measurement outcome". 

A more general definition of "realistic" (still in the context of a theory of physics) would request that the entities and objects appearing in the laws of physics correspond to entities and objects that appear in reality. Here, two items have to be made more precise:
\begin{enumerate}
\item What are the (essential) elements of the theory for which corresponding elements should exist in reality?
\\
The answer to this question becomes easier if it is possible to refer to some formal model of the theory. For example, a 
causal model  of QT may be called a realistic model, if  the elements (i.e. components) of the system state of the causal model have a corresponding element in reality.
\item  What does it mean to be an element of reality?
\\
A possible answer to this question is that an element of a model has a corresponding element in reality, if it is possible
to perform measurements for the respective elements including its components and essential attributes.
With QT/QFT, however, there exists a special problem. The laws of QT/QFT imply certain limitations with respect to the 
 possibility to measure arbitrary elements of the theory. As long as there does not exist an agreed upon  theory (i.e., "interpretation of QT") of the QT measurement process it is difficult to argue whether these limitations are (1) a consequence 
of  inherently limited measurement capabilities or rather (2) an indication that QT is not a "realistic" theory.
\footnote{The causal model of QT presented in Section 4  leads to the conclusion that it is (1), a consequence 
of  inherently limited measurement capabilities, and therefore realistic models of QT/QFT are possible.}
\end{enumerate}
The following sections of this article will focus on local causal models. The extent to which the proposed causal model of
QT/QFT described in Section 4 may be considered to represent also  a realistic model  will be discussed in Section 5.4.

\section{QT areas that cannot be directly mapped to a causal model}

In the following areas of QT, the author encountered severe problems with his attempts to construct a causal model of QT. Part of the “problem areas”, such as the measurement problem, are well-known QT problems for many years and there exist proposals for solving the problems. Nevertheless, the respective area is included here because (1) the proposed solutions apparently are not generally agreed upon by quantum physicists and/or (2) the proposed solutions do not support a mapping to a causal model. This does not necessarily exclude the feasibility of the causal models for these problem areas.  In Section 4, a causal model of QT/QFT is proposed, which includes the problem areas. 
However, assuming some causal models of QT, including the problem areas, the construction of a  \emph{local} causal
model appears to be very difficult or impossible.

\subsection{QT measurement}

The measurement problem of QT must still be considered unsolved.
The measurement problem can be expressed by a set of questions related to the overall question of what exactly happens during a measurement. The set of questions varies depending on selected basic assumptions to start with. A concise description of the measurement problem is given in \cite{Maudlin} in the form of a trilemma. In \cite{Maudlin}, Maudlin shows that the following three claims are mutually inconsistent:
\begin{enumerate}
\item The wave function of a system  is complete.
\item The wave function always evolves in accord with a linear dynamical equation (e.g., Schr\"odinger equation).
\item Measurements always have a definite outcome.
\end{enumerate}
Maudlin shows variations of these contradicting claims that are contradicting as well. 
 
Another perception of the measurement problem, which relates the measurement problem to the causal model subject of this article,  is the lack of an agreed theory of the QT measurement \emph{process}. 
QT consists of the principles, rules, and equations that describe how the probabilities (in the form of probability amplitudes) dynamically evolve in various situations to enable the prediction of the probability of different measurement results. 
The ultimate transition of the probability amplitudes to facts (i.e., to  measurement results) thus is an essential
element of QT. However, there is no agreed upon theory for this transition from probability amplitudes to facts.

Several proposals have been published under the name "interpretation of QT". Apparently, so far,  there does not exist a generally agreed upon interpretation (i.e., theory) which could be taken as the basis for a mapping to a local causal model.
Among the proposed  interpretations of QT only a few describe the measurement process in terms of a sequence 
process steps such that they enable the derivation of causal models. 
Many physicists consider decoherence theory (see \cite{Schloss}) to provide a solution to the measurement problem if it is supplemented
by the many worlds theory (see \cite{Everett}). Although the author recognizes the great contribution of decoherence theory to the progress in QT and that
decoherence theory contains elements of a causal model (i.e., a process-oriented model), he sees several open questions with decoherence theory, as well as
with the many worlds theory.

In Section 4, a causal model of QT/QFT is proposed that includes a causal model of the measurement process. Thus, a causal model of QT/QFT
that includes a measurement process (and offers a solution to the measurement problem) appears to be feasible. However, none of the proposed QT interpretations
(including the causal model proposed in Section 4) renders a mapping to a local model (according to the definition of locality given in Section 5). The
major inhibitor is the collapse of the wave function (or of an equivalent global function in the case of the many worlds theory) whose inclusion in a suitable model
of the QT measurement process appears to be unavoidable. The wave function collapse (or a similar function, such as the overall measurement), must start at a definite
position in space. Its propagation to affect the complete wave is apparently instantaneous.
\footnote{QT interpretations which do not assume a wave function collapse (such as the many-worlds interpretation, see
\cite{Everett}) assume instead some other instantaneous global process step for which it is equally difficult to construct a  local causal model.}

\subsection{Entanglement}
Entanglement is the original example in which the impossibility of a local causal model is recognized. Within QT, the entanglement of two objects (e.g., two particles) exists if a dependency among the observable attribute values (e.g., spin) of the objects exists. In the formalism of QT, the correlation of the observables can be expressed in
a common wave function, such as

(1)    $ | \Psi_{entangle} > \;\; = (1 / \sqrt{2})  (  | \uparrow >_{1} | \uparrow >_{2} +  | \downarrow >_{1} | \downarrow >_{2} ) $.
\\
Because QT is a non-deterministic theory, the dependency among the attribute values can be expressed in the relationships between the probabilities for the measurement values.
Entanglement can occur between arbitrary particle types and for different particle attributes. In the QT literature, the most frequently discussed examples of
entanglement are the spin entanglement of photons or electrons. Different types of entanglements (i.e., correlations between the measurement values) may exist.
The simplest type of correlation requests that the measurement results are equal if the same measurement type is performed for both particles. "Anticorrelation"
exists if the opposite measurement values are predicted. The following description focuses on the spin entanglement of electrons with identical predictions for the measured spin of both electrons. 
\\
\\
\begin{figure}[ht]
\center{\includegraphics*[scale=1] {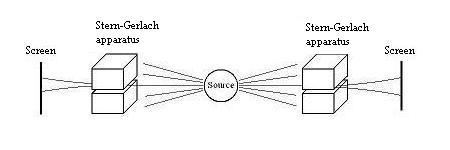} }
\caption{Components involved in the measurement of entangled electrons}
\end{figure}
Spin measurement of electrons is typically performed by the use of a Stern-Gerlach apparatus (see Fig. 1). The Stern-Gerlach apparatus can be set up with different
spatial orientations. QT predictions for the probabilities of the measurement results depend on the particle’s spin direction 
( $ particle_{i}.spindir $ ), the spatial orientation of the Stern-Gerlach apparatus ( $ SGorientation_{k}  $ ) and the expected measurement outcome. For the expected measurement outcome, only two cases
are possible: "case1" (i.e., "up") and "case2" (i.e., "down"). For the spin orientation and the spatial orientation of the 
Stern-Gerlach apparatus, only the relative orientation, i.e., the angle between the two directions, is relevant.

     $ p_{k} = P(particle_{i}.spindir-SGorientation_{k}, case1) $.
\\
Let us assume the measurement setup is such that in case of a non-entangled particle ($ particle_{i}) $,
the probability distribution for case1 measurement results is

$  P(particle_{i}.spindir-SGorientation_{1}, case1) = p_{1} $,

$ P(particle_{i}.spindir-SGorientation_{2}, case1) = p_{2} $,

...

$ P(particle_{i}.spindir-SGorientation_{n}, case1) = p_{n} $,
\\
For example, QT predicts $  p_{k} = 1 $, if $ particle_{i}.spindir-SGorientation_{k} = 0 $, 
$  p_{k} = 3/4  $, if $ particle_{i}.spindir-SGorientation_{k} = 30^0 $, or 
$  p_{k} = 1/4  $, if $ particle_{i}.spindir-SGorientation_{k} = 60^0 $. 
\\
Because there are only two alternative outcomes possible, case1 and case2, the probability  for the measurement result case2 is

$  P(particle_{i}.spindir-SGorientation_{k}, case2) = 1 - p_{k} $, if

 $  P(particle_{i}.spindir-SGorientation_{k}, case1) =  p_{k} $.
\\
Measurements on both \emph{non-entangled} particles would result in probabilities for combined measurement results which are the product of the individual result probabilities, i.e., 

(2) $ P_{combined}(particle_{1}.spindir-SGorientation_{k}, case1, particle_{2}.spindir-SGorientation_{j}, case1) = p_{j} \cdot p_{k} $. 
\\
In case of entangled particles, the relation (2)  is not true, but the probabilities for measurements on the second particle depends on the outcome of the first measurement. This is, because for entangled particles QT assigns a common probability amplitude for the combination of the two entangled particles, for example in terms of the common wave function 
(1), 

   $ | \Psi_{entangle} > \;\; = (1 / \sqrt{2})  (  | \uparrow >_{1} | \uparrow >_{2} +  | \downarrow >_{1} | \downarrow >_{2} ) $.
\\
With the chosen entanglement example (i.e., spin measurement for electrons with identical correlation) instead of equation (2) 
QT predicts 
 
(3) $ P_{entangled}(particle_{1}.spindir-SGorientation_{k}, case1, particle_{2}.spindir-SGorientation_{k}, case1) =  
   p_{k} $,

if  $  P(particle_{i}.spindir-SGorientation_{k}, case1) =  p_{k} $.
\\
Consequently, the probability for differing measurement outcomes is

(4) $ P_{entangled}(particle_{1}.spindir-SGorientation_{k}, case1, particle_{2}.spindir-SGorientation_{k}, case2) = 0 $.
\\
However, the entanglement that is predicted by QT applies not only to measurements with identical SGorientations (i.e., the orientation of the Stern-Gerlach apparatus)
but also to requesting correlated probabilities for the measurement of different SGorientations (called imperfect correlations).
\\
(5) $ P_{entangled}(particle_{1}$.spindir-SGorientation$_{k}$, case1, particle$_{2}$.spindir-
\\
SGorientation$_{j}$, case1 ) = $ px_{jk}  $.
\\
QT supports the computation of the probability amplitude and the probability $ px_{jk}  $  for entanglement with an arbitrary combination of  SGorientation$_{k} $ and SGorientation$_{j} $.

Based on assumptions that have been described as "locality, realism and causality", J. Bell derived his inequality, which requests specific relations among the $ px_{jk} $.
In \cite{Bell}, the inequality is formulated as

(6) $ | P(a,b) - P(a,c) | \leq  1 + P(b,c) $.
\\
In the notation used in this article, the inequality has to be written 
\\
$ | P_{entangled}(particle_{1}$.spindir-SGorientation$_{a}$, case1, particle$_{2}$.spindir-SGorientation$_{b}$, case1 ) - 
\\
 $ P_{entangled}(particle_{1}$.spindir-SGorientation$_{a}$, case1, particle$_{2}$.spindir-SGorientation$_{c}$, case1 ) $ |  \leq  1 +  \\
 P_{entangled}(particle_{1}$.spindir-SGorientation$_{b}$, case1, particle$_{2}$.spindir-SGorientation$_{c}$, case1 ).
\\
The probabilities that are predicted by QT  (i.e., the  P()  in (6) and $ px_{jk} $ in (5)) violate Bell's inequality. Experiments have resulted in agreement with the predictions
of QT, and QT physicists interpret this result as an indication that local causal models of QT entanglement are not possible. 

In Section 4, a causal model of
QT/QFT is proposed that includes QT entanglement. However, the proposed causal model is not a local causal model. In Sections 6 and 7, the causal model of
QT/QFT is refined such that the non-localities are confined to quantum objects.

\subsection{Interference collapse}

In textbooks of QT, interference of multiple paths of a wave function is typically explained in terms of the double-slit experiment. The double-slit experiment is also used to explain under which circumstances the interference "collapses". In  \cite{Feynman} the following phrasing is used to explain  the collapse of the interference:

"If an experiment is performed which is capable of determining whether one or another alternative is taken, the probability of the event is the sum of the probabilities for each alternative. The interference is lost."
\\
Similar phrasings explaining the double-slit experiment can be found in most basic textbooks on QT. The referenced formulation of the "interference collapse rule" cannot be mapped to a causal model because conditions such as "is capable of determining" cannot be translated to a formal causal model. In \cite{Dieldslit}, an improved interference collapse rule is suggested, which relates the interference collapse to measurements and the wave function collapse. As a consequence, a causal model of the interference collapse can be established provided a causal model of the measurement process can be given. However, the feasibility of a local causal model is questionable for the same reasons that inhibit the feasibility of a local causal model of the wave function collapse. The collapse of the interference as occurring with the double-slit experiment and the delayed-choice (Gedanken) experiment occurs apparently instantaneously.

\subsection{Quantum Field Theory}

QFT provides an extensive framework in terms of scattering matrix, Feynman diagrams, and Feynman rules for the treatment of interactions (e.g., scatterings) between particles. Nevertheless, this framework is not suitable for a mapping to a local causal model because all these powerful QFT tools, while supporting the computation of probability amplitudes for the possible final interaction results, do not enable the derivation of a continuous sequence of transformations to intermediate states. The QFT computations are based on the (idealized) assumption of a time span from $- \infty $ to $+ \infty $. In addition, Feynman diagrams must not be viewed as representing processes. Typically, the sum of multiple Feynman diagrams is required to compute the possible result probabilities for a specific type of scattering.

The above described problems (and further ones) impede the provision of a causal model of QFT, but it does not exclude the
feasibility of a causal model of QFT. In Section 4 a causal model of QT/QFT is proposed which includes a causal model of QFT interactions. 
The causal model of QFT interactions breaks down the total interaction process into a number
of more detailed process steps with related intermediate states.
However, the model described in Section 4  is a non-local (causal) model. Some state changes affect the complete 
"interaction-object". The feasibility of a local causal model of QFT interactions is questionable.

\section{A Causal Model of QT/QFT}

The proposed causal model of QT/QFT is formulated in terms of a system state that contains a discrete number of space points and discrete time intervals for the update of the system state. 
This may be implemented in the form of an extended cellular automaton (CA), as described in \cite{DielCALagr}. The extensions, 
relative to a standard CA, consist of 
\begin{enumerate}
\item an extended system state, which contains "quantum objects" (e.g., particles) and fields in addition to the cell states of the standard CA, and
\item extensions in the state-update-functions, which may access the quantum objects as a whole in addition to updating the individual cells (i.e., space points).
\end{enumerate}
Quantum objects are the entities in QT that are treated by a common single wave function and probability amplitude. These are primarily particles but also collections of particles, which must be treated as an entity.
\begin{verbatim}
System-state := {
  Space,
  Quantum-objects,    
  Fields;
}
\end{verbatim}
Space := \{ spacepoint ...\};
\\
Quantum-objects := \{ Quantumobject ...\};
\\
Quantumobject  := Particle OR ParticleCollection  OR  InteractionObject OR  
\\OtherQuantumobject;
\\ 
Fields:= $ \phi_{1}, ...,  \phi_{k} $;
\\
The most general type of quantum object is the ParticleCollection. (The InteractionObject will be used below.
The OtherQuantumobject will not be of interest within this article.)
A quantum object may be viewed as having a two-dimensional structure (see Table 1). One dimension represents the elements of the 
quantum object; the other dimension represents alternatives that may be selected during the evolution of the quantum object, for example, by an interaction. In reference to QFT (in particular, R. Feynman’s formulation of quantum electrodynamics  (QED), see \cite{Feynman1}), a quantum object consists of multiple \emph{paths}.  Each path has has an associated probability amplitude.

\small{
\begin{verbatim}
Quantumobject :=   
  path[1],
  ...
  path[n];
\end{verbatim}
}
\normalsize
\small{
\begin{verbatim}
path :=   
  pathstate[1], ...,pathstate[k], amplitude;
\end{verbatim}
}
\normalsize
\begin{table}
\caption{\label{label}Structure of a quantum object consisting of two particles p1 and p2.}
\begin{tabular} { | c | c | c  | c | }
\hline
paths & p1-state & p2-state  & amplitude  \\

\hline

path$_{1} $	  &   p1.pathstate$_{1}  $    &  p2.pathstate$_{1} $  & ampl$_{1} $  \\

path$_{2} $	 &   p1.pathstate$_{2}  $    &  p2.pathstate$_{2} $   & ampl$_{2} $  \\

...	            & ...         & ...    & ...  \\

path$_{n} $	 &  p1.pathstate$_{n} $    &  p2.pathstate$_{n} $  & ampl$_{n} $  \\
\hline
\end{tabular}
\end{table}  
The state of a path of a quantum object (the above-denoted pathstate) consists of the state-components that are known from QFT. For QED, these are, first,
the parameters that are used in QFT to specify a matrix element of the scattering matrix
$ \Psi_{p1,\sigma1,n1;p2,\sigma2,n2, ...} $,
 i.e., the four-momenta $ p^{\mu}$, the spin z-component (or for massless particles, helicity) $ \sigma $ 
and the particle type n. In addition, the position
vector x is part of the state.

At the highest level of specification are the laws of QT/QFT, which, in the formal definition of the causal model in Section 2, are subsumed in the function
\\
(7) $ applyLawsOfPhysics(state, timestep) $  :=  \{  
\\
FOR ( $ all fields \; \;  \phi_{i} $ )  \{
 
   field-state $ (  \phi_{i}$) = field-update-function($ \phi_{i}, timestep) $;
\\
\} 
\\FOR ( all quantum-objects $  qobject_{k} $ )  \{
 
  $ interactions \{  ia_{1}, ..., ia_{n} $ \} = 
 
           determine-potential-QFTinteractions( $ qobject_{k}, qobject_{j} $);
 
   IF (  NOT EMPTY($  interactions $ ) THEN  \{
 
         selected-ia = RANDOM( interactions \{  $  ia_{1}, ..., ia_{n} $ \});
 
           perform-QFT-interaction(selected-ia );
 
  \}
 
   ELSE  FOR ( all paths $  qobject_{k}.path_{i} $ )  \{
 
             $ qobject_{k}.path_{i} $ =   qobject-update-function($ qobject_{k}.path_{i}, timestep $);
 
  \}
\\
 \}
\\
 \}
\\
For a complete formal causal model, the functions  field-update-function(), qobject-update-function(), 
determine-potential-interactions(), etc. have to refined. For field-update-function() and qobject-update-function() this can be relatively easily derived from the standard QT/QFT. 
In contrast, the functions that are related to the treatment of interactions (determine-potential-interactions() and 
perform-QFT-interaction()) are non-trivial and key for the construction of the causal model for the problem areas described in Section 3. 
Therefore, field-update-function() and qobject-update-function() are not further described within this article.
The QFT interaction process is further described below. 

\subsubsection{Occurrence of an interaction} Interactions occur between two particles. If an interacting particle
is part of a quantum object (e.g., a particle collection), the effect of the interaction may propagate to other parts of the quantum object (however, this will not be addressed within this article). 

Particles are quanta of waves. In terms of wave equations (i.e., the equations of motion for the particles waves)  (see \cite{Strassler}),  an interaction between two waves $ \psi_{1} $ and $ \psi_{2} $ resulting in a third wave $ \psi_{3} $ is described by an equation of motion in which the product of waves  $ \psi_{1} $ and $ \psi_{2} $ is related to $ \psi_{3} $, as, for example, in

 $ d^{2}\psi_{3}/dt^{2} - c^{2} d^{2}\psi_{3}/dx^{2} =  a^{2} \psi_{3} + b \cdot \psi_{1} \psi_{2} $ 
\\
Many details of QFT concerning interacting particles/fields can be derived from the
interaction part of the Lagrangian. 
In  \cite{McMahon} (page 170), the Lagrangian of QED is given by 
 
 $ L_{QED} = L_{EM} + L_{Dirac} + L_{int} $
\\
with

$ L_{EM} =  - \frac{1}{4} F_{\mu \nu} F^{\mu \nu} $

$  L_{Dirac} = i \bar{ \psi } \gamma^{\mu} \partial_{\mu}  \psi  - m \bar{ \psi }  \psi  $

$  L_{int} = -q \bar{ \psi } \gamma^{\mu} \psi  A_{\mu}  $.
\\
$  L_{int} $ is the interaction part of the Lagrangian.

An interaction between particles/waves $  \psi_{1}(x,t) $ and $  \psi_{2}(x,t ) $ occurs if, for a position $ x_{0} $ the product $  \psi_{1}(x_{0},t_{0}) \cdot \psi_{2}(x_{0},t_{0} ) $ becomes non-zero, which means that both $ \psi_{1}(x_{0}) $ and $ \psi_{2}(x_{0}) $ have to be non-zero. 
The interacting particles typically consist of multiple paths. This means that the product $  \psi_{1}(x_{0},t_{0}) \cdot \psi_{2}(x_{0},t_{0} ) $ may be non-zero at multiple locations (i.e., paths) $ x_{0}, x_{1}, ..., x_{n} $. Therefore, the statement 
 
    "determine-potential-QFTinteractions(  $ qobject_{k}, qobject_{j} $)" 
\\
shown above returns in general multiple potential QFT- interactions"  \\  \{$ ia_{1}, ... ia_{n} $ \}.
\\
The term "QFT-interaction" refers to interactions whose effect is determined by the rules and techniques of QFT (e.g., scattering matrix, Feynman diagrams and Feynman rules). Scatterings are typical QFT-interactions. QFT-interactions result in further processing, as described below.
Interactions that are not "QFT-interactions" are interactions that affect multiple (or all) paths of a
particle concurrently without leading to a reduction of the path set.
(This kind of interaction might be called a "weak interaction".) For interactions that are not QFT-interactions, the further processing continues for all paths. 

After the potential QFT-interactions have been determined, only one of them is selected for processing. All other paths, i.e.,
the remaining potential interactions and the non-interacting paths, are discarded (i.e., ignored during the further processing).
This may be viewed as the reduction or "collapse of the wave function". 
The selection of the (single) QFT-interaction that will be continued is performed randomly based on the product of the
probability amplitudes of the interacting particle paths
 (see above  $  \psi_{1}(x_{0},t_{0}) \cdot \psi_{2}(x_{0},t_{0} ) $ ).
 
\subsubsection{perform-QFT-interaction()}
The second and major part of the causal model of the treatment of QFT-interactions is represented in the above formal
specification (7) in the statement "perform-QFT-interaction(selected-ia );". "selected-ia" refers to the selected paths of the interacting particles.

An overview of the quantum objects that are involved in the QFT-interaction is shown in Fig. 2. The interaction occurs between the two "in" quantum objects. The
information from the interacting paths is merged into the "interaction-object". The processing of the interaction object results in a single particle collection
(i.e., a quantum object) that contains the two "out" particles. 
(The additional particle shown at the right hand side of Fig. 2  is  particle 3 from the "in" particles which is not involved in the interaction.)
\begin{figure}[ht]
\center{\includegraphics*[scale=0.5] {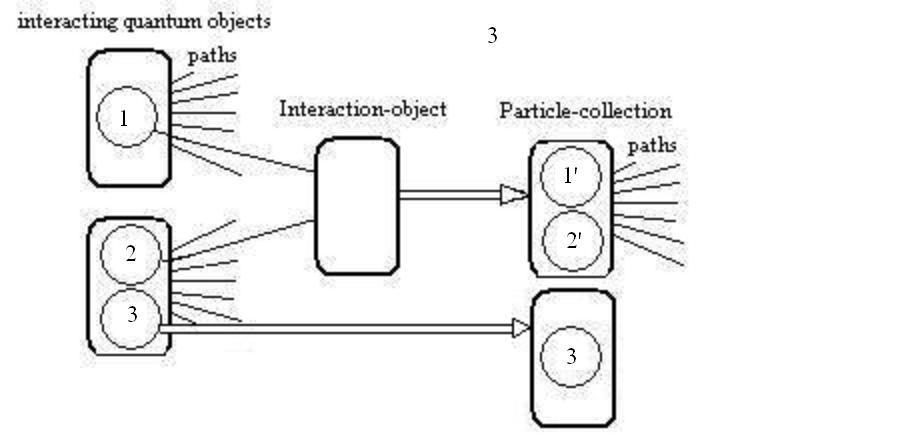} }
\caption{Quantum objects involved in QFT-interaction}
\end{figure}

The process perform-interaction() is broken down into finer more detailed process steps that are suitable for a
causal model of QFT-interactions:
\\
\\
perform-interaction( $ particle_{1}, particle_{2} $, iaposition  x) :=  \{  

   iaobject = create-interaction-object($ particle_{1}, particle_{2} $,  x);

   drop-particle( $ particle_{1}$);

   drop-particle( $ particle_{2} $);

   FOR ( all other particles in quantum object $ particle_{k} $ ) \{

   eliminate-unaffected-paths($ particle_{k} $); 

   \}

   iaresultobject =  process-interaction-object(iaobject);
\\
\}
\\
In the following these interaction process steps are roughly described. More details can be found in  \cite{Dielfi} and 
 \cite{DielCALagr}.
\begin{itemize}
\item create-interaction-object($ particle_{1}, particle_{2} $,  x);
\\
The information from the interacting particles $ particle_{1}$ and  $ particle_{2} $  is combined into an "interaction object". The interaction object is a special type of quantum object.
\item drop-particle( particle);
\\
The particle is destroyed.
\item eliminate-unaffected-paths($ particle_{k} $); 
\\
All of the paths of $ particle_{k} $, except for the path that caused the interaction, are discarded. Elimination of all of the paths, except for the path that
caused the interaction, supports the causal model of entanglement, as described in Section 4.4.
\item process-interaction-object(iaobject);
\\
The processing of the interaction object is determined by the rules and equations of QFT (e.g., Feynman diagrams, Feynman rules, Fermion chains).
However, these QFT rules must be mapped to a causal model. Details of this mapping are described in  \cite{Dielfi} and   \cite{DielCALagr}. The overall result of the QFT-interaction
is embraced in a \emph{single} particle collection (i.e., a quantum object), which is herein called the iaresultobject. The iaresultobject typically contains two particles (which can be of the same type as that of the ingoing particles) and multiple paths. The generation of a single particle collection ensures proper correlations with the alternative outcomes of the interaction.
\end{itemize}

\subsection{Causal Model of Interactions in Quantum Field Theory}

QFT-interactions are the focus area with the proposed causal model of QT/QFT for two reasons:
\begin{itemize}
\item QFT-interactions are the major subject within QFT that causes problems with respect to the provision of a local causal model.
For the remaining areas of QFT, the mapping to a causal model is relatively straightforward.
\item QFT-interactions are the major basis from which a causal model of the other problem areas
(measurement, interference collapse, entanglement) can be derived.
\end{itemize}
It is therefore  worthwhile to summarize the main characteristics of the above causal model relative to
QFT-interactions.
\begin{enumerate}
\item QFT-interactions occur at a definite point in space.
\item Because the interacting particles/waves in general occupy larger areas of space, there may be multiple points in space that are candidates for the interaction location. QT/QFT randomly selects one of the candidate locations as the single interaction position. The random selection is based on the probability amplitudes of the candidate paths. This ensures that the probability distribution (and differential cross sections) of QFT-interactions is in accordance with the predictions of QT/QFT.
\item After a single path of (both) the interacting particles/waves has been determined to be the interacting path, the remaining paths of the interacting \emph{particle collections} become obsolete. For the causal model, this may be interpreted either as (1) the obsolete paths being ignored during the further processing, (2) the obsolete paths being discarded ($ \rightarrow $
collapse of the wave (function)), or (3) the "obsolete" paths branching into new worlds ($ \rightarrow $ many worlds interpretation ).
The author claims that all three cases represent a discontinuity in the linear progression according to the wave equation (e.g., Schr\"odinger equation). The causal model of QT/QFT described above assumes alternative (2), i.e., the elimination of obsolete paths.
\footnote{However, this choice is not essential for the overall causal model.} 
\item The interaction result depends only on the interacting paths of the particles.
\item In general, the interaction result consists of a particle collection (i.e., a quantum object) consisting of two particles with multiple (alternative) paths. Each path has an associated (single) probability amplitude.
\item The interaction result (i.e., the particle collection) may be considered as a mapping of the "in" particles to the "out" particle collection. This mapping, however, does not represent a bijective mapping.
\footnote{This is not in contrast with the unitarity of the scattering matrix.} 
\end{enumerate}

\subsection{Causal Model of Interference Collapse}

In \cite{Dieldslit}, an improved interference collapse rule is proposed as follows:
\\
"Interference is lost if the particle/wave becomes involved in a QFT-interaction on its way to the observation target."
\\ Involvement in a "QFT-interaction" refers to 
\\
"determine-potential-QFTinteractions()" and "perform-interaction()" in the above
causal model of QT/QFT. Particularly, the elimination of unaffected paths (see "drop-particle()" and  "eliminateunaffected-paths( )") is responsible for the interference collapse.

\subsection{Causal Model of QT measurement}

The above described causal model of QT/QFT does not contain a dedicated special part for QT measurement. 
The causal model of QT measurement  is based on the assumption that the measurement process can be mapped to "normal"
QT/QFT  functions that are part of the overall causal model.

For the mapping of a measurement process to normal QT/QFT functions, QFT-interactions play a major role.
Measurements of QT parameters can be performed using a variety of measurement devices, apparatuses and processes. Common to all such measurement processes is that they have to include at least one interaction in which the measured object exchanges information with some other object belonging to the measurement apparatus. 
QFT-interactions, as described above, are the only type of interactions suitable for the information exchange required for QT measurements. Nevertheless, this does not mean that QFT-interactions enable arbitrary information exchange (i.e., measurements). The following types of limitations of QFT-interactions are mainly responsible for the peculiarities of QT measurement: 
\begin{itemize}
\item Because a QFT-interaction always starts with the selection of a single location/path
\footnote{If multiple paths meet at a common position, this leads to interference that affects the probabilities of possible measurement results.}
for which the interaction is performed, all information that is only represented by the totality of the wave cannot be measured. This causes the uncertainty of QT observables and transition to definite measurement results.
\item For some specific particle/wave attributes, the measurement information cannot (directly) be retrieved. For example, the spin direction of a particle cannot be directly measured, but there exist only interactions that support the measurement of 
 the projection of the spin to the direction of the measurement apparatus (i.e., $ \sigma \cdot  \vec{J} $ ).
\item A single QFT-interaction can only provide information regarding a single particle attribute (e.g., position). This means that it is not possible to measure multiple attributes concurrently. The solution used for classical measurements, namely, using multiple successive interactions to obtain the information regarding multiple attributes, does not generally help because each QFT-interaction destroys or modifies most of the  information.
\item QFT-interactions support only a non-bijective mapping of the "in" state to the "out" state and, thus, only a limited exchange of information. This limited exchange of information is the cause of some of the limitations and peculiarities of QT measurements.
\end{itemize}

\subsection{Causal Model of Entanglement}

Entanglement experiments are special cases of QT measurements. Therefore, the entire description above for QT measurement also applies to the measurement
of entanglement. However, there exist additional entanglement-specific aspects that must be considered for the establishment of a causal model of entanglement.
Entanglement can occur with arbitrary particle types and for different particle attributes. As with the entanglement considerations in Section 3.2, the
following description focuses on the most frequently discussed type of entanglement, the spin entanglement of electrons. The proposed causal model of entanglement
refers to the spin entanglement but with minor modifications is also applicable to other types of entanglement.
For the construction of a causal model of entanglement,
it is necessary to consider the complete process, starting with the creation of the entangled pair of particles until the determination of the correlated measurement
results. Fig. 1 shows the components that are involved in the measurement of entangled particles. The following process steps are distinguished:
\begin{enumerate}
\item At the source: Creation of the entangled pair of particles
\\
The creation of the entangled pair of particles is the result of a QFT-interaction and therefore creates a particle collection,
as described in Section 4.
\begin{table}
\caption{\label{label}Particle collection consisting of two entangled particles .}
\begin{tabular} { | c | c | c  | c | }
\hline
paths & particle1-state & particle2-state  & amplitude  \\

\hline

path$_{1} $	  &   particle1.pathstate$_{1}  $    &  particle2.pathstate$_{1} $  & ampl$_{1} $  \\

path$_{2} $	 &   particle1.pathstate$_{2}  $    &  particle2.pathstate$_{2} $   & ampl$_{2} $  \\

...	            & ...         & ...    & ...  \\

path$_{n} $	 &  particle1.pathstate$_{n} $    &  particle2.pathstate$_{n} $  & ampl$_{n} $  \\
\hline
\end{tabular}
\end{table}   
\item Propagation of paths in different directions
\item At the Stern-Gerlach apparatus: Path diversion as a function of spin orientation and the orientation of the Stern-Gerlach apparatus (see Section 3.2; $ P(particle_{i}.spindir-SGorientation_{k}, case1) $). The path diversion is accompanied by 
an adjustment of the particles spin direction ($ particle_{1}.spindir = SGorientation_{k} $). This adjustment causes a corresponding adjustment of the spin direction of the entangled particle ($ particle_{2}.spindir $).
\item At the screen: Measurement of the particle positions 
\\
This "measurement" implies a typical QFT-interaction, as described in Section 4.1, with the reduction of the path set of the particle collection to the single path that covers the interaction position. The reduction of the path set applies to both of the entangled particles.
\end{enumerate}

\section{\emph{Local} Causal Models}

\subsection{Locality}

The definition of a local causal model presupposes a spatially causal model (see Section 2.2). A (spatially) causal model is understood to be a local model if
changes in the state of the system depend on the local state only and affect the local state only. The local state changes can propagate to neighboring locations.
The propagation of the state changes to distant locations; however, they must always be accomplished through a series of state changes to neighboring locations.  
\footnote{Special relativity requests that the series of state changes does not occur with a speed which is faster than the speed of light. This requirement is not considered within the present article.}

Based on a formal model definition (such as the formal causal model defined in Section 2), a formal definition of locality can be given. We are given a physical theory
and a related spatially causal model with position coordinates x and position neighborhood dx  (or 
$ x  \pm \Delta x $ in case of discrete space-points).

A causal model is called a local causal model if each of the laws $ L_{i} $ applies to no more than a single
position x and/or to the neighborhood of this position $ x  \pm dx $.
\\
In the simplest case, this arrangement means that $ L_{i} $ has the form

  $     L_{i} : \: IF \:c_{i}(s(x)) \: THEN  s'(x) = f_{i}(s(x)); $
\\
The position reference can be explicit (for example, with the above simple case example) or implicit by reference to a state component that has
a well-defined position in space. References to the complete space of a spatially extended object are considered to violate locality. References to specific properties
of spatially extended objects do not violate locality.

The above definition of locality,
where locality strictly refers to space-points (i.e., x and   $ x  \pm dx $) is called in this article
"space-point locality". To enable the construction of causal models, which are not space-point local but should not beclassified as completely non-local, a weaker form of locality, called “object-locality”, is described in Section 5.3.
\\
Note: The above definition of locality does not impose any limitations on the speed by which objects can move or on the speed of inter-object communication.
If Lorentz invariance is a further desired property of the (model of the) theory (which is typically the case), this property is considered to be a separate, additional requirement
that is not addressed within this paper. For the above given definition of locality, it is only requested that $ \Delta t $  (the state update time interval with the
invocation of the physics-engine) is greater than 0; thus, 
 $  \Delta x /  \Delta t < \infty $.

\subsubsection{Example3 - A local causal model} A specific CA that represents a local causal model can be specified for the evolution of a wave.

The standard formulation of the "wave equation" (i.e., the equation of motion for waves) is (see, e.g., \cite{Chun})

(3)  $ ( \frac{1}{v^{2}} \frac{d^{2}}{dt^{2}} - \frac{d^{2}}{dx^{2}}) \psi (x,t) = 0 $.
\\
To obtain the CA update-function, the equation of motion has to be transformed into a sequence of computation steps, including the replacement of the differential operations by discrete "$ \Delta $-units". The following computation steps for the CA  are derived from the above wave equation for the state transition $ \psi(x_{i},t_{j})  \rightarrow \psi(x_{i},t_{j+1}) $
\begin{enumerate}
\item $ t_{j+1} = t_{j} + \Delta t $
\item $  \Delta \psi dx = ( \psi(x_{i+1},t_{j}) - \psi(x_{i-1},t_{j}) ) / 2  \Delta x $
\item $ \Delta^{ 2} \psi dx  = ( \psi(x_{i+1},t_{j}) - 2  \psi(x_{i},t_{j}) + \psi(x_{i-1},t_{j})) / \Delta x /  \Delta x $
\item wave equation: $ \Delta^{ 2} \psi dt =  v^{2}  \Delta^{ 2} \psi dx  $
\item $ \Delta^{ 2} \psi dt =  ( \psi(x_{i},t_{j+1}) - 2  \psi(x_{i},t_{j}) + \psi(x_{i},t_{j-1})) / \Delta t /  \Delta t 
\rightarrow  $
\\
$  \psi(x_{i},t_{j+1}) = \Delta^{ 2} \psi dt \cdot \Delta t \cdot  \Delta t +  2  \psi(x_{i},t_{j}) - \psi(x_{i},t_{j-1}) $
\end{enumerate}  
(The following naming conventions are used for going from the differential units to $ \Delta$ -units: $ d \psi/dt \rightarrow \Delta \psi dt $, $ d \psi/dx \rightarrow \Delta \psi dx $,   $ d^{2} \psi / dx^{2}  \rightarrow  \Delta^{ 2} \psi dx $, $ d^{2} \psi / dt^{2}  \rightarrow  \Delta^{ 2} \psi dt $).

\subsubsection{Example4 - A non-local causal model:}

In \cite{French}, page 120, a coupled oscillator that consists of two pendulums coupled through a spring is described (see Fig. 3).
Assume a system (state) that contains the two masses Ma and Mb and the related parameters.
\\
\begin{figure}[ht]
\center{\includegraphics*[scale=0.5] {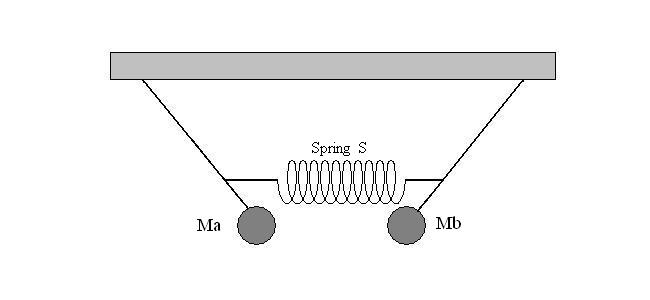} }
\caption{Two coupled pendulums}
\end{figure}

System-state := \{
 
  Space;
   
  Masses  Ma, Mb;

  Ma := $ \{ m_{a}, x_{a}, \omega_{a}, initx_{a} \}$; 

  Mb := $ \{ m_{b}, x_{b}, \omega_{b}, initx_{b} \}$;
\\
\}
\\
The meanings of the attributes are as follows: masses  $ m_{a} $ and $ m_{b}$;  pendulum displacements  $ x_{a}$ and $ x_{b}$; frequencies  $ \omega_{a} $ and $  \omega_{b}$; and  initial pendulum displacements $ initx_{a} $ and $ initx_{b}$.
The equations of motion for the two special cases (1) $ initx_{a} = initx_{b} $ and (2) $ initx_{a} = - initx_{b} $
can be expressed by

update-function :=

  $     L_{1} : \: IF ( initx_{a} = initx_{b} ) \; \;  THEN \; \;  \\
  \{  \omega' = ( \omega_{a}^{2} + 2 k / m_{a})^{1/2}; \; \; x_{a} =C cos \;  \omega' \;  t; \; \; x_{b} =C cos \; \omega' \;  t  \}$ 

  $     L_{2} : \: IF ( initx_{a} = -  initx_{b} ) \; \;  THEN \; \; \\
  \{  \omega' = ( \omega_{a}^{2} + 2 k / m_{a})^{1/2};  \; \; x_{a} =C cos\; \omega' \;  t; \; \; x_{b} = -C cos \; \omega' \;  t  \}$  
\\
The appearance of both $  x_{a} $ and  $ x_{b} $ within the laws $   L_{1},  L_{2} $ establishes the non-locality. 

With this specific example, the non-locality can be eliminated with a refinement of the example causal model such that the pendulum displacements  $  x_{a} $ and  $ x_{b} $
are derived from a common cause for the two pendulums and/or by showing more  coupling details through the explicit propagation of state changes.

\subsection{Non-localities within the proposed causal model of QT}

The proposed causal model of QT/QFT violates locality in each of the four problem areas:
\begin{itemize}
\item QFT-interaction 
\\
The collapse of a particle’s wave function (which, in the causal model, is expressed
by the reduction to a single path) represents a non-local action. Moreover,
the mechanism, which ensures that only a single interaction occurs if multiple candidate interaction positions
exist, represents a non-locality. The
processing of the interaction object (see Section 4) is not modeled at a level
of detail that enables the determination of possible further non-localities. However,
the author doubts that a more detailed mapping of the QFT mechanisms
(e.g., Feynman diagrams, Feynman rules) to a causal model can be accomplished
such that non-localities can be avoided. Instead, it may be
feasible to identify QFT-interactions as the source of all of the non-localities
within QT.

\item Interference Collapse
\\
Because  the  interference collapse is a consequence of a QFT-interaction,  it implies the same non-localities as 
QFT-interactions in general (see above).
\item QT Measurement
\\
Because  a  QT measurement implies at least one QFT-interaction,  it implies the same non-localities as QFT-interactions in general (see above).
\item Entanglement
\\
The causal model of entanglement contains QT measurements and thus QFT-interactions. Therefore, entanglement
implies the same non-localities as QFT-interactions in general (see above).
Moreover, the causal model of
entanglement contains the type of non-locality that leads to the violation of
Bell's inequality. The apparent concurrent update of both entangled particles
during a measurement (see Section 4.4) represents the most prominent
non-locality within QT.
\end{itemize}
The listed non-localities in the causal model of QT/QFT are first of all violations of space-point locality. It may be feasible to transform the space-point non-localities to a model that supports object locality. However, this requires non-trivial modifications of the causal model of QT/QFT described in Section 4. In Section 6, a refinement of the causal model of QT/QFT is described where the non-localities are confined to quantum objects.

\subsection{Object locality}

According to the above given definition of a (space-point) local causal model, the causal model of QT/QFT described in Section 4 contains several non-localities. To enable the construction of reasonable causal models where space-point locality cannot generally be provided, but the non-localities can be confined to well-defined sub-units of the overall system state, “object locality” is defined as follows:

A causal model is called an \emph{object-local} causal model, if  the system state of the causal model contains compound spatially extended objects and all references within the laws  $ L_{i} $ of the causal model are either space-point local references or references to the global properties of an object.

\subsection{Is the proposed causal model a realistic model?}

Quantum physicists consider Bell's famous inequality and its violation in experiments
to be a strong indication that "local realistic" models of QT are not possible.
In 2.3, a definition of a realistic model is given that requests that the entities and
objects that appear in the model correspond to entities and objects that appear in
reality. As a refinement, the phrase "appear in reality" is understood to be "can be measured".

It is certainly a major goal of any physicist who develops a theory of physics
to formulate the laws of the theory in terms of objects and parameters that
can be measured. This formulation includes entities that cannot be measured directly but
can be computed only as a function of other (measurable) entities. This goal
of using as many (directly or indirectly) measurable parameters as possible may
be assumed to hold for the formulation of QT and the formulation of the
causal model of QT/QFT. Nevertheless, as is well-known, there exist severe limitations
with respect to the measurability of QT observables. The search for a
possible explanation for these limitations is part of the QT measurement problem.
The causal model of QT measurement described in Section 4.3 explains the
peculiarities and limitations of QT measurements by the special characteristics
of the QFT-interactions, which are a mandatory ingredient of all QT measurements.
Thus, the difficulties in mapping the elements of the theory to the elements of reality
are not caused by the fact that the elements of the theory are "unrealistic"
but rather by the limitations of the mapping (i.e., measurement) capabilities.

\section{Refined Causal Model with non-localities confined to quantum objects}

The conclusion of Section 5.2 that the causal model of QT/QFT contains non-localities
supports the conjecture that local causal models of QT and of QFT are
not possible. Because it is hardly imaginable that QT in general and in principal
is a non-local theory, two questions arise:
\begin{enumerate}
\item Is it possible to remove some of the non-localities by a refined
and more detailed causal model of QT/QFT?
\item Is it possible to confine the apparently non-removable non-localities to specific areas or aspects of QT/QFT? 
\end{enumerate}
It is indeed possible to observe that the non-localities in the proposed causal
model appear whenever compound entities within the system state, such as
particle collections or single particles that consist of multiple paths, must be
treated as an entity. A second observation is that the areas and processes in which
the non-localities appear are all related to quantum field theory and, in particular,
to QFT-interactions. Within this article, the compound entities (i.e., particle
collections, including single particles) that occur in the causal model of QT/QFT
have been called quantum objects. It is therefore reasonable to attempt a refinement
of the causal model such that the non-localities are confined to quantum
objects and quantum objects are studied in more detail.

\subsection{Refined overall causal model}

In Section 2, a causal model is defined as being embedded in a "physics-engine".
In Section 4, the proposed causal model of QT/QFT is described by a system state
that contains quantum objects and by an execution logic of the physics-engine,
which at the highest level starts with a loop around all quantum objects, i.e.,
\begin{verbatim}
FOR ( all quantum-objects  qobject[k] ) {
........
}
\end{verbatim}
According to Section 5.1, the reference to spatially extended objects or to a set
of objects must already be considered to be a violation of locality. (In a truly local
causal model, the highest level loop would refer to space-points.) An elimination
of this high level non-locality can be achieved by the assumption that a single central physics-engine that knows all quantum objects
and controls their concurrent dynamical evolution does not exist, but each quantum object
is completely autonomous (i.e., runs its own copy of the physics-engine, see
Fig. 4). This view is also compatible with Special Relativity, in which the global time
and the "proper time" of the local system are distinguished. The local proper time
of a quantum object is achieved by the local physics-engines with individual
speeds in their update cycles.
\\
\begin{figure}[ht]
\center{\includegraphics*[scale=0.5] {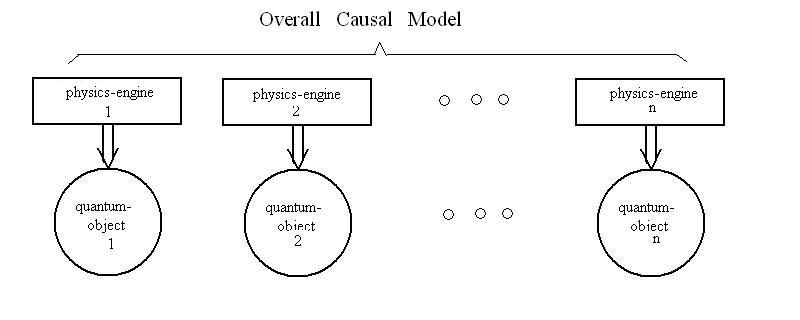} }
\caption{Causal model with separate physics-engine per quantum object}
\end{figure}
Assuming that quantum objects autonomously run in parallel and have their own
physics-engine does not mean that the physics-engines (which represent the
laws of physics) are required to be different. Differences in the dynamical evolution of
the quantum objects must be caused purely by the differences in the quantum
object-local system state. Moreover, the dynamical evolution of the quantum
objects should depend on only the quantum object-local
system state as much as possible.
\footnote{Otherwise it may not be possible  to claim even object-locality}

The quantum object’s local physics-engine represents the laws of physics
that are applicable to the progression of the individual quantum object. There
are, however, further laws of physics that define the overall interrelations, such
as, for example, how the objects move in space and when they will meet at common
space-time points. For the (refined) causal model of QT/QFT, it is assumed
that synchronization between quantum objects is required and can occur with
interactions only. Rather than assuming some overall physics-engine that
would be responsible for the inter-object relationships, the causal model assumes
that the overall synchronizations, such as the determination of when two quantum
objects meet and interact, is also the task of the individual quantum objects
(i.e., their physics-engines). This arrangement means that the physics-engines of the
autonomous quantum objects must determine the space-time-points of interactions
between the quantum objects in accordance with the equations of motion.
This construct appears to be feasible only in a causal model in which the movements of the particles are controlled (i.e., determined and kept track)in terms of global or invariant quantities for the position in space and velocity.
In other words, the synchronization of autonomous quantum objects
that move in different trajectories is accomplished through the common 
space. 

Another global subject that must be mapped to object local treatment
is the global conservation of quantities such as the energy, momentum, and angular momentum.
With the causal model of QT/QFT, this goal is achieved by the bookkeeping
of these quantities with each interaction (see below).

\section{Quantum objects}

Quantum objects are collections of particles (including single particles) whose collective dynamical evolution and measurement results can only be described by the laws of QT/QFT. 
\\
\\
The collective dynamical evolution and the implications for measurement results typically are expressed by a common wave function. An example is the wave function for two spin-entangled particles $   \Psi_{1} $ and $  \Psi_{2} $

  $ | \Psi_{entangle} > \;\; = (1 / \sqrt{2})  (  | \uparrow >_{1} | \uparrow >_{2} +  | \downarrow >_{1} | \downarrow >_{2} ) $
\\
which specifies the quantum object  $  \Psi_{entangle} $.
\\
There are further examples for how in QT the collective dynamical evolution of multiple (virtual) particles is specified. 
For example, the equation of QFT, which defines the possible scattering processes in quantum electrodynamics (QED), specifies also (if supplemented by the rules of QFT) a special quantum object.

 $ H_{W}(x) = -eN \{ ( \bar{\psi^{+} } +  \bar{\psi^{-} }) ( \not A^{+} + \not A^{-}) ( \psi^{+} -  \psi^{-}) \}_{x} $
\\
$ \bar{ \psi^{+}}, \bar{ \psi^{-}}, \not A^{+}, \not A^{-},  \psi^{+},  \psi^{-} $ are creation and annihilation operators ( see \cite{Mandl}, page 111).  The mapping of this equation (and further laws of QFT) to a causal model of QFT resulted in a type of quantum object called “interaction-object”  in Section 4.

In summary, the following examples of quantum objects occurred within this article:
\begin{enumerate}
\item Single particles  (measurement, interference collapse)
\item Interaction object  (QFT interaction)
\item Collection of particles resulting from QFT interaction  (measurement)
\item Entangled particles  (measurement)
\end{enumerate}
In parenthesis the related non-local functions (as described in Section 5.2) are listed.
There are further examples of quantum objects, such as bound systems which, however will not be further discussed
within this article.

\subsection{Causal model of the quantum object}

\subsubsection{The state of the quantum object}

The assumption that quantum objects run autonomously leads to the objective that the state upon which the (physics-engine of the) quantum object operates should be as much as possible local to the quantum object. The refined
causal model of QT/QFT assumes the following quantum object state.
\\
\\
\begin{figure}[ht]
\center{\includegraphics*[scale=0.7] {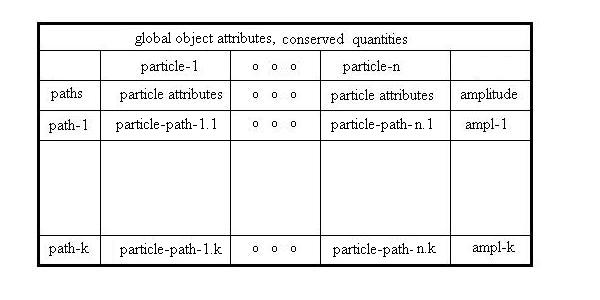} }
\caption{Structure of the Quantum Object}
\end{figure}
Quantum-object-state := \{
 
  global-attributes;

   conserved-quantities;

  particle-set := \{ $ particle_{1}, ..., particle_{k}$  \}; 
\\
\}
\\
$ particle_{i}  :=  \{ type, mass, path_{1}, ..., path_{j}$  \};
\\
$ path_{i} := \{ amplitude,  pathstate_{1}, ..., pathstate_{k} $  \}; 
\\
pathstate := \{ spacepoints, momentum, angularmomentum \};
\\
spacepoints := \{ $  sp_{1}, ...,  sp_{n} $ \}; 
\\
conserved-quantities := \{energy, momentum, angularmomentum \};
\\
global-attributes := \{position, momentum \};
\\
The global parameters "global-attributes" and "conserved-quantities" (see Fig. 5) serve for two purposes 
\begin{enumerate}
\item They enable the necessary coordination between two interacting quantum objects. 
With the creation of a new quantum object, the global parameters are derived from the corresponding global parameters  of two interacting quantum objects.
After the creation of the interaction object the conserved-quantities remain static until possible further interactions with a 
quantum object occur.
\item They serve as a  base for the path-related equivalent quantities. The  path-related equivalent quantities (spacepoints, momentum, angularmomentum ) are relative to the global parameters.
\end{enumerate}
The dependency of the quantum object dynamics on the global parameters position and conserved-quantities
may be viewed as representing non-localities.  However, if this view is taken, it would represent a "non-locality" which is 
confined to the quantum object. As will be shown below, the assumption of the  global parameters enables the
construction of causal models where the otherwise unspecified kind of non-localities can be confined to the
quantum object level. 

\subsubsection{Dynamics  of the quantum object}

The causal model of QT/QFT described in Section 4 has to be slightly adapted to 
cover only object local dynamical evolution and to access only object local information.
\\
(8) $ applyLawsOfPhysics(state) $  :=  {  
\\FOR ( all particles  pa$_{k} $ )  \{
 
   interactions \{ $ ia_{1}, ..., ia_{n} $ \} = 
 
           determine-potential-QFTinteractions( $ pa_{k}, externalparticle_{j} $);
 
   IF ( NOT EMPTY( interactions )) THEN  \{
 
           selected-ia = RANDOM( interactions \{ $ ia_{1}, ..., ia_{n} $ \});
 
           perform-QFT-interaction( $ pa_{k}, externalparticle_{j} $, iaposition );
 
  \}
\\
 \}
\\FOR ( all particles  pa$_{k} $ )  \{

 $ pa_{k}.path_{i} $ =   qobject-update-function($ pa_{k}.path_{i} $);
\\
 \}
\\
Except for the elimination of the timestep parameter there are only minor changes to the overall causal model of QT/QFT
described in Section 4. The more significant changes with respect to the locality of the referenced information occur with the details of the function  perform-QFT-interaction( ). In Section 4 the function  perform-QFT-interaction( ) is specified as follows:
\\
\\
perform-interaction( $ particle_{1}, particle_{2} $, iaposition  x) :=  \{  

   iaobject = create-interaction-object($ particle_{1}, particle_{2} $,  x);

   drop-particle( $ particle_{1}$);

   drop-particle( $ particle_{2} $);

   FOR ( all other particles in quantum object $ particle_{k} $ ) \{

   eliminate-unaffected-paths($ particle_{k} $); 

   \}

   iaresultobject =  process-interaction-object(iaobject);
\\
\}
\\
The detailed functions have to be adapted as follows:
\begin{itemize}
\item create-interaction-object($ particle_{1}, particle_{2} $,  x);
\\
For the creation of the interaction object only the information from the two particles at their common position x plus 
quantum-object-global parameters are required. The interaction object is assumed to be created at position x. Thus,
the function may be considered space-point-local.
The new created interaction object is a quantum object which gets associated a new (copy of the) physics-engine
and a complete quantum object state. For the initialization of the "conserved quantities" of the new quantum object 
the respective quantities from the two interacting particles ( $ particle_{1}$ and $ particle_{2} $) have to be 
summarized and adjusted by the values at position x.

This creation of the new quantum object called interaction object means that the further processing of the 
interaction object described below continues autonomously in parallel with the further processing (i.e., termination) of the two 
interacting quantum objects.
\item drop-particle( particle);
\\
The instantaneous disappearance of a complete  spatially extended particle may be considered as a non-local function.
The apparent non-locality can be circumvented by assuming a dependency of the continued evolution of the particle on a particle global parameter such as the particles mass. 
\item eliminate-unaffected-paths($ particle_{k} $); 
\\
Similar to drop-particle (above), the instantaneous disappearance of almost the complete  spatially extended particle may be considered as a non-local function.
The elimination of the unaffected paths of the partner particle (of the interacting particle) may be implemented by making the further progression of the unaffected paths dependent on some global attribute of the quantum object.
\item process-interaction-object(iaobject);
\\
As described above ( see "create-interaction-object") the processing of the interaction-object 
continues autonomously in parallel with the termination of the two 
interacting quantum objects. 
\end{itemize}
 

\section{Discussions}

\subsection{To what extent are local causal models of areas of physics worthwhile?}

The assumptions that J. Bell made when he derived his inequalities have been described as locality, realism, and causality. 
Many physicists  apparently agree with, for example, T.F. Jordan writing in \cite{Jordan} (p. 125) "Bell inequalities are obtained from ideas about objective reality and causality that appear to be good common sense."  This explains the astonishment of physicists when the inequalities were violated in experiments. 
There are numerous further examples in physics where the physical laws are described in terms of processes or 
(Gedanken-) experiments which may be viewed as representing a kind of causal model. Nevertheless, this does not mean that there exist in general complete and coherent causal models for physics theories. The descriptions preferred by physicists for the specification of a physical theory are "declarative descriptions" which state the laws of physics in terms of equations (mostly differential equations) for the relationships between the parameters and properties (i.e., state components) of the physical system.
With classical theories of physics it is in general easy and straightforward to derive the related causal model. In general, 
the equations of motion which can be derived from the (declarative) Lagrangian  are sufficient to derive 
a causal model.

With QT severe  doubts may be raised as to whether a causal model that can be derived from the declarative equations of QT such as the  Schr\"odinger equation  can be sufficiently complete (see the problematic QT areas described in Section 3).
Some (not the majority of) physicists seem to be dissatisfied with the lack of a causal model of QFT. For example, R. Feynman wrote in \cite{Feynman1}:
\\
 "I have pointed out these things because the more you see how strangely nature behaves, the harder it is to make a model that explains how even the simplest phenomena actually work. So theoretical physics has given up on that." 
\\
Explaining "how phenomena actually work" is equated with the provision of a causal model in this article.

The usefulness of local causal models of areas of physics does not mean that the model has to be local and causal according to the (somewhat strong) definitions given in Section 2. If it turns out that for major areas of physics, such as QT/QFT, the construction of a local causal model according to the definitions given in Section 2 is not possible, it is necessary to search for other, possibly weaker, types of locality and/or causality for which it is possible to construct a consistent and complete model of subject theory.

\subsection{The completeness of QT}

As mentioned in Section 1 Introduction, A. Einstein questioning the completeness of QT lead to the EPR experiment and to
Bell's inequality. Some physicists therefore consider the violation of Bell's inequality in experiments as a refutation of Einsteins doubts for the completeness of QT. 
Others think that the incompleteness of QT does not consist (as Einstein suspected) in the fact that QT makes wrong predictions in case of entangled particles, but  that the fact that apparently it is not possible to construct a local causal model of entanglement can be interpreted as an incompleteness of QT.
In \cite{DielComplete} the author argues that a theory of physics for which it is not (yet) possible to construct a complete
causal model should be called incomplete.


\section{Conclusion}

Causal models of quantum theory are feasible if certain conditions are satisfied:
\begin{enumerate}
\item The term "local causal model" requires a more precise definition. The most precise
and most useful definition of a local causal model is a definition that is made in terms of a \emph{formal}
model.
\item The lack of a well-defined, generally agreed-upon theory in certain areas of
QT currently prevents the construction of a complete causal model of QT.
The construction of a causal model for the problem areas enables a complete
causal model and, at the same time, implies a proposal for the QT completion
in the problem areas.
\item The proposed causal model of QT/QFT (see Section 4) demonstrates
that causal models of QT/QFT are feasible. However, the causal model of
QT/QFT proposed in Section 4 contains non-localities. 
\end{enumerate}
Concerning \emph{local} causal models of QT/QFT, it is concluded that
\begin{enumerate}
\item Local causal models of QT/QFT are not possible if the locality is understood to be the “space-point locality”.
\item A refined causal model of QT/QFT in which the non-localities are confined to “quantum objects” is described in Sections 6 and 7.
 The resulting kind of locality is called object-locality. Object-locality is obtained by the introduction of compound objects
(here quantum objects) and the assignment of global properties to the object.
\item The establishment of quantum objects (and of compound objects, in general) together with the identification of object-global properties is not just a technical trick that supports the construction of object-local causal models.
The resulting object-local causal model may have significant physical implications leading to new explanations for existing problems and/or the disclosure of new questions.
\end{enumerate}

%

\end{document}